\documentclass[pre,twocolumn,superscriptaddress,preprintnumbers,amsmath,amssymb]{revtex4-1}

\usepackage{epsfig}
\usepackage{amsmath}
\usepackage{amssymb}
\usepackage{amsfonts}
\usepackage{mathptmx}
\usepackage{dcolumn}
\usepackage{eucal}
\usepackage{bm,bbm}
\usepackage{color}
\usepackage[colorlinks,linkcolor=blue,citecolor=blue]{hyperref}
\usepackage{epstopdf}
\usepackage{bbold}
\usepackage{url}

\usepackage{dsfont}

\newcommand{\DetectorName}{{PSRD}}

\begin{document}

\title{Proximity SQUID single photon detector via temperature-to-voltage conversion} 

\author{P. Solinas}
%\email[]{paolo.solinas@spin.cnr.it}
\affiliation{SPIN-CNR, Via Dodecaneso 33, 16146 Genova, Italy.}
\author{F. Giazotto}
\affiliation{NEST, Instituto Nanoscienze-CNR and Scuola Normale Superiore, I-56127 Pisa, Italy}
\author{G. P. Pepe}
\affiliation{Universit\'a di Napoli Federico II, Monte Sant'Angelo, I-80125 Napoli, Italy}
\affiliation{SPIN-CNR, Via Dodecaneso 33, 16146 Genova, Italy.}

\date{\today}

\begin{abstract}
We propose a single photon detector based on a superconducting quantum interference device (SQUID) with superconductor-normal metal-superconductor Josephson weak links.
One of the two Josephson junctions is connected to an antenna, and is heated when a photon is absorbed.
The increase of the weak link temperature exponentially suppresses the Josephson critical current thereby inducing an asymmetry in the SQUID. This generates a voltage pulse across the SQUID that can be measured with a threshold detector.
Realized with realistic parameters the device can be used as a single photon detector, and as a calorimeter since it is able to discriminate photons frequency above $5~$THz with a signal-to-noise ratio larger than $20$.
The detector performance are robust with respect to working temperatures between $0.1~$ K and $0.5~$K, and thermal noise perturbation.
\end{abstract}

\date{\today}

\maketitle

\section{Introduction}
\label{sec:intro}

%%%%%%%%%%%%%%%%%%%%%%%%
Photon-counting detectors demonstrating the highest sensitivity and efficiency have become indispensable for many applications in the visible and near infrared electromagnetic range such as space-to-ground communications \cite{ Moision2008} or quantum key distribution \cite{Gisin2002}.
The benchmark for solid-state single photon detector is determined by the avalanche photodiodes but in $2001$ a novel superconducting single-photon optical detector was proposed \cite{Goltsman2001}.
The main advantages of this device is to be sensitive at visible and infrared wavelengths, work at relatively high temperature, i.e., $4.2~$K and be more sensitive 
of the Si avalanche photodiode in the long wavelength region \cite{Natarajan2012}.
In the last decade there has been an increased interest followed by a great technological development \cite{Natarajan2012}. 
This includes an improve in coupling efficiency using nanowire meander \cite{Miki2008}, in absorption efficiency using nanowire in optical cavities \cite{Rosfjord2006, Miki2010} and in registering efficiency using ultra-narrow nanowire (of $20$ or $30~$nm) \cite{Marsili2011}.
Fabrication with material alternative to the original NbN has been proposed and tested \cite{Miki2009, Shibata2010, Curtz2010}.
The same devices are promising candidates for the detection of multi-photon events \cite{Dauler2007,Dauler2009}.

The need for high sensitivity detectors also in the THz region is of particular interest in many research fields ranging from communication technology, passing to astrophysics to arrive to the more exotic implication in quantum information. 
Far-infrared (far-IR) including the THz region is a very important spectral range since it contains about 98\% of all the photons existing in the Universe: the availability of very efficient recording devices at long-wavelength single photon regime in this spectral range represents  a truly exciting frontier in astronomy \cite{Echternach2017}. 
%Most of the energy coming from newly forming stars and accreting black holes manifests in the far-IR \cite{Echternach2017}, and similarly, cooling of optically thick disks of planets assembling. 

However, single photon THz detectors are significantly limited by the difficulties to obtain a single-photon sensitivity due to  the small energy carried by an individual photon (i.e. typically 2-5 meV). 
This energy corresponds to a detector equivalent temperature of only a few tens of kelvins, and hence extremely low operating temperatures are needed to avoid the thermal noise.
Besides the first detection of single THz photons with semiconductor quantum dots (QDs), hot electron and superconducting bolometers \cite{Komiyama2000, Wei2008, Tarkhov2006}, new and efficient detection schemes are needed.
The possibility of using the physical properties of the superconducting state in Josephson-based devices represents an attractive degree of freedom to be explored due to their high charge and phase sensitivity in determining final  performances. 

%The possibility to detect the arrival of a photon with frequency in the terahertz regime is of great importance in both research and technology.
%The possible applications range from communication technology, passing to astrophysics to arrive to the more exotic implication in quantum information.
%In situation in which the detector can be cooled at low temperature, superconducting devices represent a viable alternative because of the high sensibility.

Here, we propose an original design for  a proximity SQUID radiation detector (PSRD) that can be used as a single photon sensor and as a calorimeter, i.e., by measuring the frequency of the incident photon.
A proximized normal metal in a long Josephson weak link can be heated by the absorption of the photon.
The following increase in its electronic temperature induces an instability in the interferometer that generates a voltage pulse across the SQUID in order to relax to a stable state.. The key point in the process is the possibility to set the SQUID close to an instability point.
In the present case, this can be done by adjusting the external magnetic flux piercing the interferometer.

This working principle is directly related to the observation that when we change the magnetic flux in a Josephson interferometer and cross such an instability point, the superconducting phase must undergo a phase jump \cite{giazotto2013coherent, Solinas2015JRC}.
This new physical phenomenon was originally discussed in Ref. \cite{giazotto2013coherent} and indirectly observed in Ref. \cite{giazotto2012josephson}.
The extension to the full dynamical model has been done in Refs. \cite{Solinas2015JRC,Solinas2015JRCExtended, Bosisio2015}.
In these later papers, the system was driven through an instability point by an external time-dependent magnetic flux.
The control of the time-dependent magnetic field allows one to induce the phase jump and the corresponding voltage pulse at will.
When subject to a periodic drive, the interferometer emits a voltage pulse comb (a sequence of equally spaced and identical voltage  pulses) in complete analogy to the optical combs used in metrology \cite{udem2002optical}.

The same idea is at the basis of the present \DetectorName.
We consider two Josephson weak links (JWLs) in a superconducting quantum interference device (SQUID) configuration \cite{tinkham2012introduction}.
By tuning the magnetic flux piercing the SQUID, we can set the device close to an instability point.
When a photon is absorbed (through a suitable antenna coupled to one of the JWLs) it heats the JWL inducing a phase jump, and a voltage pulse.
The detection of these voltage pulses gives us information about the photon arrival.
When the detection electronics is able to distinguish additional information about the voltage pulse, e.g., its maximum, it is possible to discriminate the photon frequency, and use the \DetectorName~as a calorimeter.

From our analysis, a Nb-based PSRD can work well for photon frequencies above $5~$THz.
With a detector bandwidth of $10~$GHz, the achievable signal-to-noise ratio and resolving power are large enough to discriminate both the photon arrival and its frequency.
The PSRD can operate at sufficiently high temperature, i.e., up to $\sim 0.5~$K of bath temperature, and it is robust against thermal noise.

%The principle of the \DetectorName~can be explained by the tilted washboard potential model with the help of figure (\ref{fig:Washboard_potential}).
%For simplicity we can consider the symmetric junctions configuration, i.e., $r=0$ in Eq. (\ref{eq:I_J}), where $I_J(\varphi; \phi) = I_+ \cos \phi \sin \varphi$.
%The Josephson energy associated is 
%\begin{equation}
% E_J(t) = \int dt I_J~V = \alpha \int d\varphi I_J = - \alpha I_+ \cos \phi \cos \varphi.
% \label{eq:E_J_symm}
%\end{equation}
%The phase particles is usually in a potential minimum [see Fig. (\ref{fig:Washboard_potential})], say, $\varphi = 0$.
%When we modulate the magnetic field we change the energy potential.
%For $ \Phi/\Phi_0 = 1/2$ (correspondind to $\phi = \pi/2$), the energy potential vanishes.
%As soon as the system crosses this critical point and $\phi = \Phi/\Phi_0 = 1/2+\epsilon$ (with $\epsilon>0$), the phase particle that is still in $\varphi = 0$ but this position correspond now to a maximum.
%Since its position is highly unstable it will roll to one of the nearest minimum at $\pm \pi$.
%This sudden jump in the phase is, by the Josephson relation, associated to a voltage pulse acrogiazotto2006opportunitiesss the junction.
%

\section{Single Josephson weak link under photon absorption}

First, we consider a single JWL consisting of a normal metal (N) wire of length $l$ coupled to two superconductors S via transparent contacts  \cite{giazotto2008Ultrasensitive, Voutilainen2010}.
The JWL is coupled to an antenna that is able to collect the incident photon radiation at energy $h\nu$ [see Fig. \ref{fig:detector_fig}a)].
The latter heats the electrons in the normal metal region thereby increasing its electronic temperature $T_e$.

We denote with $\Delta$ and $D$, the gap of the superconductor $S$ and the diffusion coefficient of the normal metal, respectively.
If the Josephson junction is long, i.e., for $E_{Th}  = \hbar D/ l^2 \ll k_B T_e \ll \Delta$ where $E_{Th}$ is the Thouless energy, the Josephson current of the junction is $I_J (T_e) = I_c (T_e) \sin \varphi$ 
where $\varphi$ is the superconducting phase difference across the junction and \cite{Zaikin1981, Wilhelm1997, giazotto2008Ultrasensitive}
\begin{equation}
  I_c (T_e) = \frac{64 \pi k_B T_e}{ (3+ 2 \sqrt{2} e R_N)} \sqrt{\frac{ 2 \pi k_B T_e}{E_{Th}}} e^{-\sqrt{\frac{ 2 \pi k_B T_e}{E_{Th}}}}.
  \label{eq:I_c}
\end{equation}
Above, $R_N = \rho l/\mathcal{A}$ is the normal-state resistance of the wire  ($\rho$ and $\mathcal{A}$ are the wire resistivity and cross section, respectively). 
In the following, we have used $D= 0.01~$m$^2$s$^{-1}$, $\rho = (\nu_F e^2 D)^{-1}$ with $\nu_F = 10^{47}~$J$^{-1}$m$^3$ is the density of states at the Fermi level in the N region and 
\cite{giazotto2008Ultrasensitive}.
The geometric parameters $l$ and $\mathcal{A}$ changes from junction to junction and are chosen to maximize the device performances (see discussion below) but, as reference, we can have 
(typical of silver, Ag), $L=1~\mu$m, and $\mathcal{A} = 10^{-15}~$m$^{2}$ leading to $R_N=38~$Ohm.

To determine the dynamics of the junction temperature we follow Ref. \cite{giazotto2008Ultrasensitive}.
We assume that a photon of frequency $\nu$ is absorbed at time $t_0$ in the volume $\Omega$ of the N region.
We model the energy injection into the junction with a Gaussian pulse with standard deviation (in time) $\sigma$ so that the photon power is $P_\gamma = \frac{2 \pi \hbar \nu}{ \Omega \sqrt{2 \pi} \sigma} e^{-(t-t_0)^2/(2 \sigma^2)}$.
The N region of the junction has an electronic thermal capacity $C_e = \pi^2 \nu_F k_B^2 T_e/3$, and it is in contact with the phonon bath so that it dissipate power (per unit of volume) as $P_{e-ph} = \Sigma (T_e^5 -T_{bath}^5)$ where $\Sigma$ is the electron-phonon coupling constant in N \cite{giazotto2006opportunities}.
The power balance equation therefore reads \cite{giazotto2008Ultrasensitive}
\begin{equation}
 C_e \frac{d T_e}{dt} =P_\gamma - P_{e-ph}.
 \label{eq:power_balance}
\end{equation}
In the following we consider volume of $\Omega = 10^{-21}~$m$^3$ and $\Sigma = 0.5\times10^{9}~$Wm$^3$K$^{-5}$ typical for a silver junction.

An example of the behavior of $T_e(t)$ is shown in Fig. \ref{fig:RCSJ_dyn} a).
The photon heats the junction up to a maximum of roughly $T_{e,{\rm max}} = \sqrt{T^2_{bath} + 12 \hbar \nu/(\pi \Omega \nu_F k_B^2)}$, then the junction dissipate the energy excess over a timescale of the order of $\tau_{\rm e-ph} $ \cite{giazotto2008Ultrasensitive}.
Since, in general, the  diffusion time along the N strip $\tau_D =l^2/D \approx \sigma  \ll \tau_{\rm e-ph}$ \cite{giazotto2006opportunities, giazotto2008Ultrasensitive} where $\tau_{\rm e-ph}$ is the electron-phonon
interaction time ($\tau_{\rm e-ph}\sim 10^{-4}\ldots 10^{-7}$s in the $0.1\ldots 1$K temperature range), we can assume that the temperature of the junction remains almost constant after photon absorption, as shown  in Fig. \ref{fig:RCSJ_dyn} a).
If we define the relaxation time $\tau_R$ as the time needed for $T_e (\tau_R) = (T_{e,{\rm max}}-T_{bath})/2$, with the parameter used for the numerics in Fig. \ref{fig:RCSJ_dyn}, we obtain $\tau_R \approx 120~$ns. That is much longer than $\tau_D\sim 0.1~$ns and $\sigma$ (for the numerical simulation we have taken $\sigma = 0.1~$ns).
In the present model, the relaxation time is the limiting factor for the photon detection rate.
However, using different materials and different fabrication geometry, the energy dissipation to the phonon bath can be increased and, thus, the idle time can be reduced.
%However, with an accurate optimization of the geometry, junction and substrate materials can be considerably reduced.

The mechanism responsible for the presence of a supercurrent in the SNS junction is the \emph{proximity} effect that induces in the N region a local density of states with a minigap $E_g \approx 3.1 ~E_{Th}$ \cite{Zhou1998, giazotto2008Ultrasensitive}. In all our analysis 
we ignore the effect that such a modification induces in
 the heat capacity and the electron-phonon
coupling \cite{Rabani2008, Rabani2009, Heikkila2009}. Specifically, thanks to the presence of the minigap, both these quantities
are expected to be somewhat suppressed inside the N region thereby improving the PSRD
performance.
Moreover, a photon with energy $\nu$ can be absorbed in the junction if $\nu > 2 E_{Th} /h$ \cite{virtanen2017josephson, heikkila2017thermoelectric}.
For the typical junction parameters used in our calculations we have $\nu > 3~$GHz.

In the present analysis we assume that when a photon is absorbed all its energy is transferred to the junction generating the increase in temperature. This description is, of course, oversimplified.
The absorption and the efficiency in the energy transfer depends on the details of the antenna, e.g., materials and geometry.
The antenna can be designed and optimized for to work on a kind of photons properties or energy region.
In the present paper, we do not deal with such details of the specific implementation and focus more on the on the physical mechanism underlying the photon detector.

\section{Dynamics of the SQUID}

\begin{figure}
    \begin{center}
    \includegraphics[scale=.75]{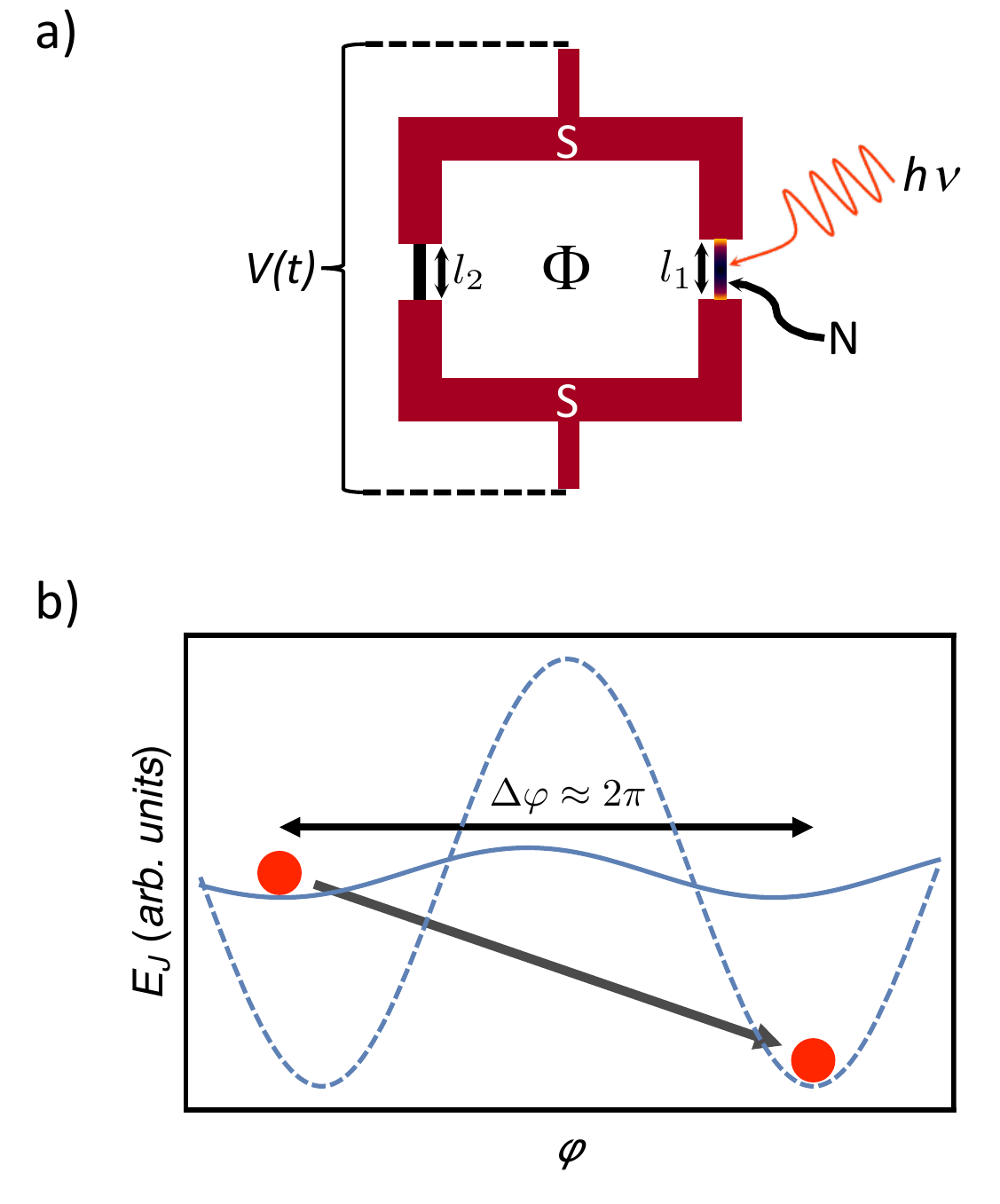}
       \end{center}
    \caption{
	a) Scheme of the PSRD. Two SNS Josephson weak links are arranged in a SQUID configuration.
	One junction is connected to an antenna that absorbs the arriving photons at energy $h\nu$. 
	The weak links lengths are $l_1$ and $l_2$, respectively, and the SQUID is pierced by an external magnetic flux $\Phi$.
	The photon absorption generates a time-dependent voltage pulse $V(t)$ across the interferometer.
	b) Pictorial representation of the induced instability. The Josephson energy $E_J$ (in arbitrary units) is plotted before (solid blue curve) and after photon absorption (dashed blue curve).
	The change in the energy potential induces a dynamics in the phase particle (red circle) that jumps to a close minimum with a phase change $\Delta \varphi$ of about $2 \pi$.
     }  
    \label{fig:detector_fig}
\end{figure}

To exploit the phase jump effect discussed in Sec. \ref{sec:intro} and \cite{Solinas2015JRC,Solinas2015JRCExtended, Bosisio2015} ,  we consider two SNS Josephson weak links in a SQUID configuration.
One of the JWLs, say junction $1$, is coupled to an antenna and it is heated by the photon, as shown in Fig. \ref{fig:detector_fig} a).
The JWLs are supposed to have the same resistance $R_{N,1} = R_{N,2}= R_N$ but different length, i.e., $l_1 \neq l_2$, in order to have different Thouless energy.
Moreover, the junction length asymmetry is captured by the  parameter $\epsilon = l_1/l_2$.
Here, we assume that the ring inductance is negligible but a similar approach can be used in case of a non-vanishing ring inductance \cite{Guarcello2017, Guarcello2018}.

\begin{figure}
    \begin{center}
    \includegraphics[scale=.6]{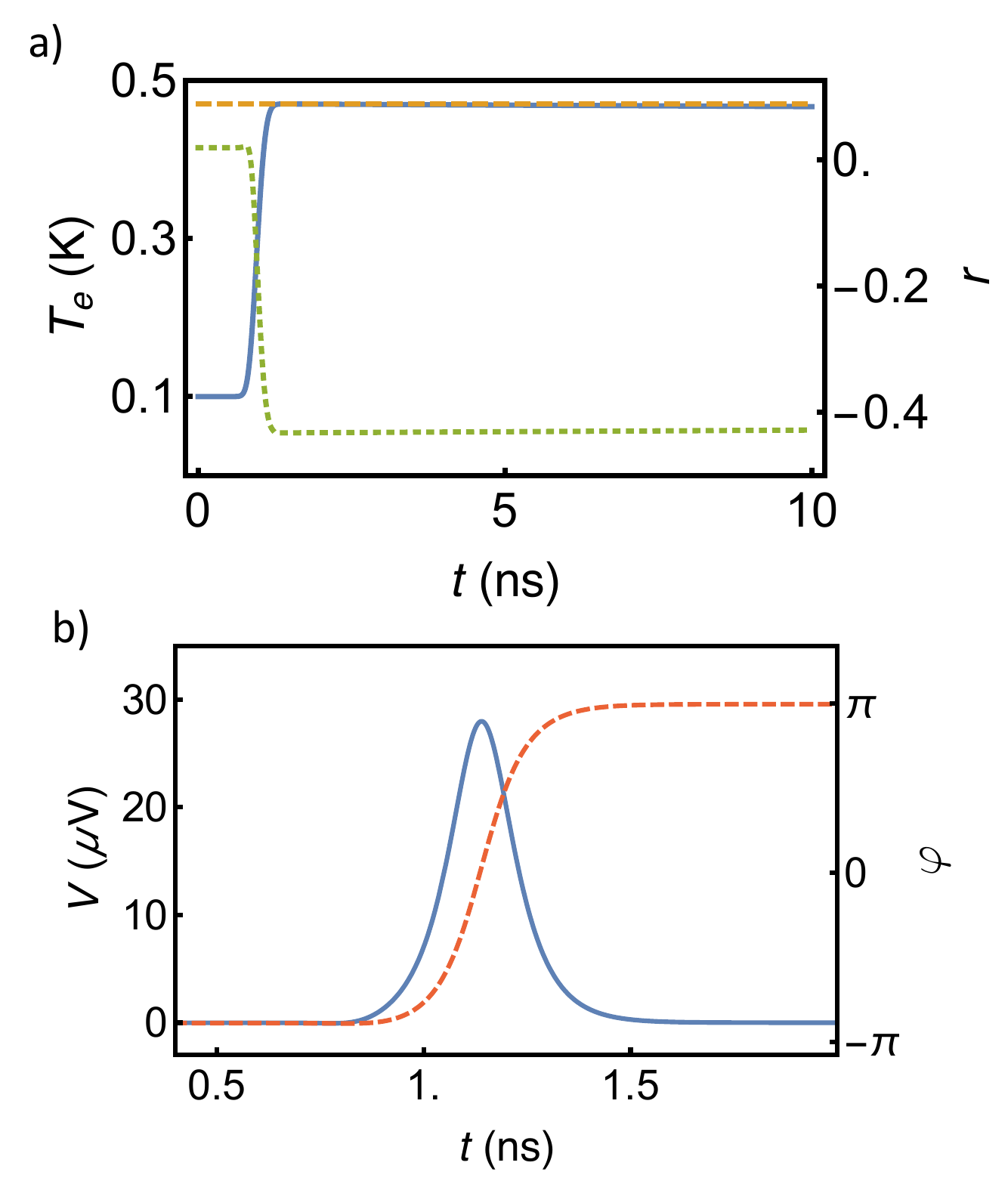}
       \end{center}
    \caption{
    %{\bf Temperature dynamics and critical current dynamics.} 
    a) Temporal dynamics of the temperature $T_e$ of the normal metal region when a photon is absorbed (solid blue curve), and of the SQUID asymmetry parameter $r$ (dotted green curve). The maximal temperature reached by the N region is showed as a dashed orange line. 
b) Voltage (solid blue) and superconducting phase (dashed red) as a function of time due to photon absorption.
    The parameters used in the calculations are  \cite{giazotto2008Ultrasensitive}: $T_{bath}=0.1~$K, $\Omega = 10^{-21}~$m$^3$, 
    $\nu_F = 10^{47}~$J$^{-1}$m$^3$ and $\Sigma = 0.5\times10^{9}~$Wm$^3$K$^{-5}$ [as appropriate for silver (Ag)], $D = 0.01~$m$^2$s$^{-1}$, $l = 10^{-7}$m, $R_N=38~$Ohm. The standard deviation of the Gaussian photon envelope is $\sigma = 0.1~$ns.
     }  
    \label{fig:RCSJ_dyn}
\end{figure} 

We denote with $\Phi$ the magnetic flux piercing the SQUID, and the current ($I_J$) vs phase relation of the SQUID reads \cite{Solinas2015JRC}
\begin{equation}
I_J(\varphi; \phi) = I_+ [\cos \phi \sin \varphi + r~ \sin \phi \cos \varphi], \,
\label{eq:I_J}
\end{equation}
%where $\varphi=(\varphi_1+\varphi_2)/2$,  $\phi = \pi \Phi/\Phi_0$ ($\Phi_0\simeq 2 \times 10^{-15}$ Wb is the flux quantum), $I_+ = I_{c1} + I_{c2}$, $\varphi_i$, $r= ( I_{c1} - I_{c2})/ I_+ $,  and $I_{ci}$ ($i=1,2$) are the phase across and the critical current of the $i$-th junction, respectively.
where $\varphi_i$ is the phase across the $i$-th junction, $\varphi=(\varphi_1+\varphi_2)/2$ and $\phi = \pi \Phi/\Phi_0$ ($\Phi_0\simeq 2 \times 10^{-15}$ Wb is the flux quantum) is the normalized magnetic flux.
If  $I_{c,i}$ ($i=1,2$) is the critical current of the $i$-th junction, we have 
\begin{eqnarray}
	I_+ &=& I_{c,1} (T_e) + I_{c,2} (T_{bath}), \\ \nonumber
	r &=& \frac{  I_{c,1} (T_e) - I_{c,2} (T_{bath})}{ I_{c,1} (T_e) + I_{c,2} (T_{bath})}.
\end{eqnarray}
To write Eq. (\ref{eq:I_J}) in this way, it is important to assume that the Josephson weak links have the same resistance $R_N$ but different critical currents, i.e., $ I_{c,1} \neq  I_{c,2}$.
The choice to keep $R_N$ fixed is arbitrary since the only physical important parameter is the asymmetry of the junction critical currents needed to increase the photon absorption effect. 
The main advantage is that it allows us to to simplify the analytical and numerical treatment.
Similar results can be obtained if we assume different junction resistances.
Experimentally, to have the same resistance the junction should be fabricated in a particular way.
A different length in the weak links determines a difference in the Thousless energy (that scales as $l^2$).
This difference length must be balanced by different section areas since we want equal resistance in the junctions.

%To write the above equation is important to assume that the Josephson weak links have the same resistance $R_N$ but different critical currents, i.e., $ I_{c,1} \neq  I_{c,2}$.
%This can be accomplished by fabrication. A different length in the weak links determines a difference in the Thouless energy (that scales as $1/ l^2$).
%This difference length must be balanced by different weak links cross sections since we want equal resistance in the junctions. [SPIEGARE MEGLIO PERCHE' FACCIAMO QUESTA SCELTA]

\begin{figure*}
  \includegraphics[scale=.65]{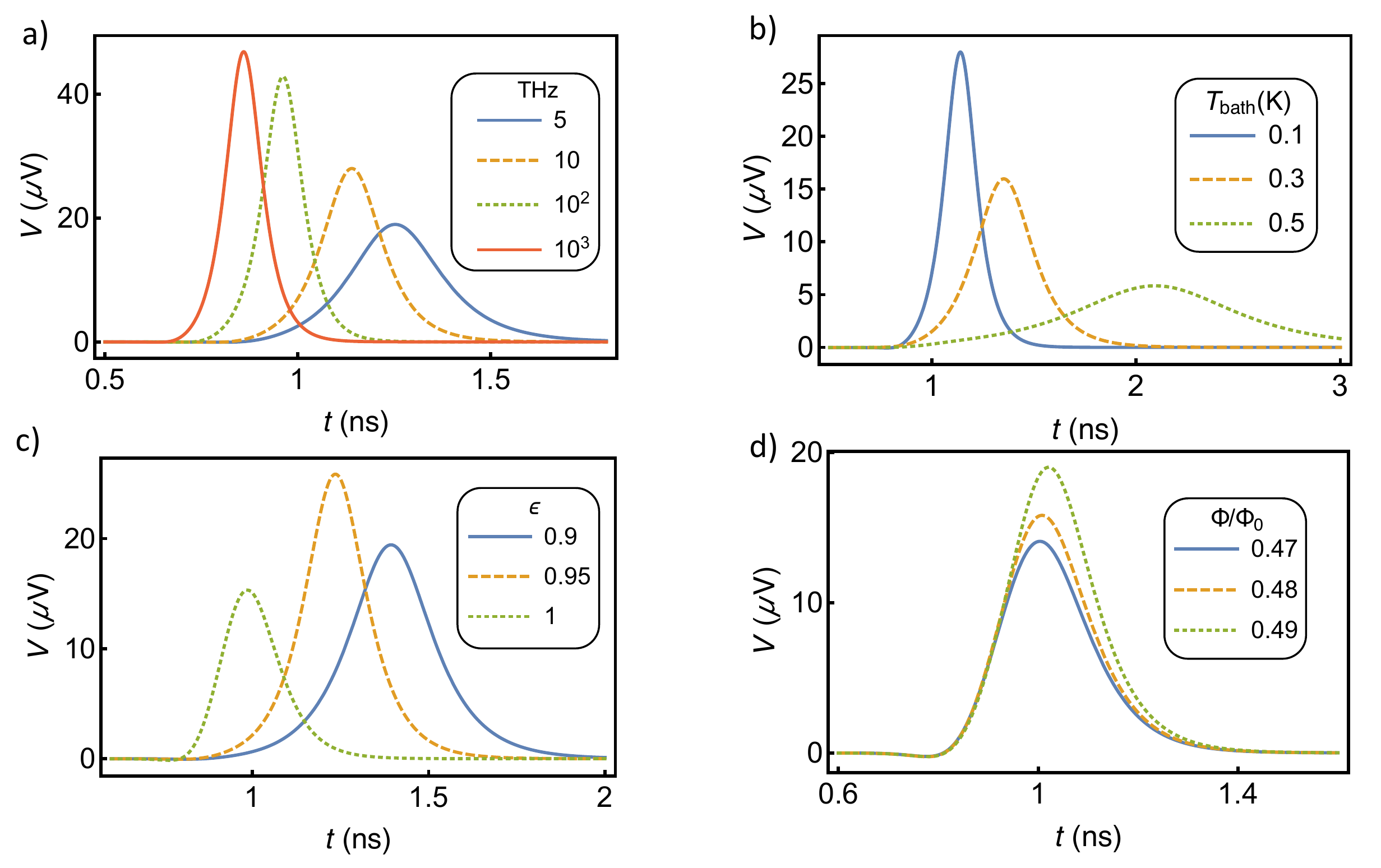}
    \caption{ 
     %{\bf Time-depedent voltage} 
     a) Voltage pulse dynamics for different absorbed photon frequencies $\nu$ at $T_{bath}=100~$mk, $\Phi/\Phi_0 = 0.499$, and $\epsilon = 0.98$.
     b) Voltage pulse dynamics for different bath temperatures $T_{bath}$ at $\nu=10~$THz, $\Phi/\Phi_0 = 0.499$, and $\epsilon = 0.98$.
     c) Voltage pulse dynamics for different asymmetry parameters $\epsilon$ at $T_{bath}=100~$mk, $\nu=10~$THz, and $\Phi/\Phi_0 = 0.499$.
     d) Voltage pulse dynamics for different magnetic fluxes $\Phi$ at $T_{bath}=100~$mk, $\nu=10~$THz, and $\epsilon = 0.98$.
     }  
    \label{fig:FigTot}
\end{figure*}

Under these hypothesis, the dynamics of the phase and the applied voltage $V$ is given by the resistively and capacitively shunted Josephson junction (RCSJ) model \cite{tinkham2012introduction, Solinas2015JRC}.  We model the SQUID as a capacitor $C$, a resistor $R$, and a non-linear, magnetic flux-dependent inductance $L_J$ arranged in a parallel configuration.
The equation for $\varphi$ can be written in terms of the dimensionless variable $\tau = 2 \pi \nu t$ as \cite{tinkham2012introduction}
\begin{equation}
   c \frac{d^2 \varphi}{d \tau^2}+  \frac{d \varphi}{d \tau} + \alpha[ f(\varphi,\tau) - \delta]=0,
  \label{eq:RCSJ_adim}
\end{equation}
where $ \delta=I_B/I_+$, $c =2 \pi R C \nu$, $f(\varphi,\tau) = I_J[\varphi;\phi(\tau)] / I_+$ and $\alpha= I_+ R/(\Phi_0 \nu)$.
In the following, we consider the case in which the junction capacitances are negligible, and there is no bias current flowing through the SQUID, i.e., $c=0$ and $\delta =0$.

Before solving numerically Eq. (\ref{eq:RCSJ_adim}), it is useful to have a qualitative understanding of the effect induced by photon absorption.
This can be done by analyzing the behaviour of the Josephson energy  as a function of the magnetic flux $\Phi$, and the time-dependent critical current $I_{c,1}$.
The Josephson energy reads \cite{tinkham2012introduction, Solinas2015JRC}
%[see Eq. (\ref{eq:E_J_symm})]
\begin{equation}
   E_J  =  \alpha \int d\varphi I_J=- \alpha \Big[I_{c,1} \cos \Big( \varphi +  \frac{\pi \Phi}{\Phi_0} \Big) + I_{c,2} \cos \Big(\varphi-  \frac{\pi\Phi}{\Phi_0} \Big) \Big].
   \label{eq:E_J_general}
\end{equation}

Let us set consider the case in which the external magnetic flux is $\Phi/\Phi_0 = 1/2 - \chi/\pi$ with $\chi \ll 1$, i.e., close to an instability point \cite{Solinas2015JRC}.

For sake of discussion, let us discuss first the case in which $I_{c,1} (t=0) \gg I_{c,2}$.
In this case, the Josephson energy reads $E_J/E_{J,0} \approx - \cos \Big(\varphi + \pi  \frac{\Phi}{\Phi_0} \Big)  =- \sin (\varphi - \chi )$ that has minimum at $\varphi =  (4m+1) \pi /2   - \chi$ (with $m$ integer).

When the photon is absorbed, $I_{c,1}$ decreases exponentially and becomes much smaller than $I_{c,2}$.
The  Josephson energy reads $E_J/E_{J,0} \approx -\cos \Big(\varphi- \pi  \frac{\Phi}{\Phi_0} \Big) = -\cos \Big(\varphi + \chi - \frac{\pi}{2} \Big) = -\sin (\varphi + \chi )$ that has minimum at $\varphi = (4k+1) \pi/2  +\chi$ (with $k$ integer).
The two energy minima are shifted by $\Delta \varphi =2 \pi (m-k)  - 2 \chi $ and  the closest {\it non-trivial} jump, i.e., $m\neq k$, is obtained for $\Delta \varphi \approx 2 \pi$.

In the realistic implementation we have $I_{c,1} (t=0) \approx I_{c,2}$ [see Fig. \ref{fig:RCSJ_dyn} a) where $r(t=0)\approx 0$].
This is represented in Fig. \ref{fig:detector_fig} b) where we pictorially show how the change in the Josephson potential energy can induce the phase jump.
It is important to notice that this energetic discussion does not give information about the amplitude of the jump (that can be  $|\Delta \varphi| = 2 \chi, 2 \pi  - 2 \chi, 4 \pi  - 2 \chi, ...$). This must be determined by solving the full dynamics of the junction temperature [Eqs. (\ref{eq:power_balance})] and phase dynamics [Eq. (\ref{eq:RCSJ_adim})].
An example of the dynamical behaviour of  $T_e$, $I_{c,1}$, $r$, $\varphi$ and $V(t)$ is shown in Fig. \ref{fig:RCSJ_dyn} a) and b).

Notice that while in the original paper \cite{Solinas2015JRC}, the phase undergoes a $\pi$ jump, here the jump is of $2 \pi$.
This difference is related to the nature of perturbation of the Josephson energy landscape.
While in Ref. \cite{Solinas2015JRC}, the time-dependent magnetic field induces the Josephson energy modulation, here it is the asymmetry of two junctions induced by the photon absorption.

%Summarizing, the absorbed photon induces effectively induces a deformation in the Josephson potential. 
%To minimize the energy the phase particle jumps from the initial minimum to the new one.
%This phase changes is related, through the Josephson relation, to a voltage pulse appearing at the extremes of the devices.
%By measuring the latter, we can determine the photon arrival and its frequency.

\section{Voltage dependence on junction and drive parameters}

The voltage generated across  the SQUID vs time is shown in Fig. \ref{fig:FigTot} as a function of  different parameters: the photon frequency $\nu$ in Fig. \ref{fig:FigTot} a), the bath temperature $T_{bath}$ in Fig. \ref{fig:FigTot} b), the asymmetry parameter $\epsilon$ in Fig. \ref{fig:FigTot} c), and the applied magnetic flux $\Phi$ in Fig. \ref{fig:FigTot} d).

Figure \ref{fig:FigTot} a) shows that for a photon frequency above $5~$THz the SQUID generates a sizable voltage pulse at $T_{bath}=0.1~$K.
The voltage pulse maximum increases with the photon frequency  ranging from $20~\mu$V at $\nu=5~$THz to $40~\mu$V at $\nu=100~$THz.

Another relevant parameter is the working temperature $T_{bath}$ [show in Fig.  \ref{fig:FigTot} b) for $\nu = 10~$THz].
At low temperature ($T_{bath}=0.1~$K), the temperature jump in the junction is large and leads to an increased asymmetry and large voltage pulses (solid blue curve).
At higher temperature  ($T_{bath}=0.5~$K), the voltage pulse is broadened with a smaller maximum around $5~\mu$V meaning that the device can work even at these temperatures.

The PSRD works at best for a small asymmetry parameter $\epsilon$, as shown in Fig.  \ref{fig:FigTot} c).
This is because the asymmetry induced by the photon absorption is maximized.
A possible asymmetry range is $0.9 \leq \epsilon \leq 1$.

The magnetic flux $\Phi$ must be close to a critical point in order to put the SQUID in a unstable point, i.e., $\Phi/\Phi_0 = 0.5$.
The best performance are obtained for $\Phi/\Phi_0 = 0.499$ but the voltage pulse is sizable as well for $\Phi/\Phi_0 = 0.47$.
However, in deciding the optimal applied magnetic flux we must also consider the effect of the noise (see Sec. \ref{sec:thermal_noise} below).
As a matter of fact, if the system is too close to an instability point, the noise could induce an undesired transition.

\begin{figure}
  \includegraphics[scale=.6]{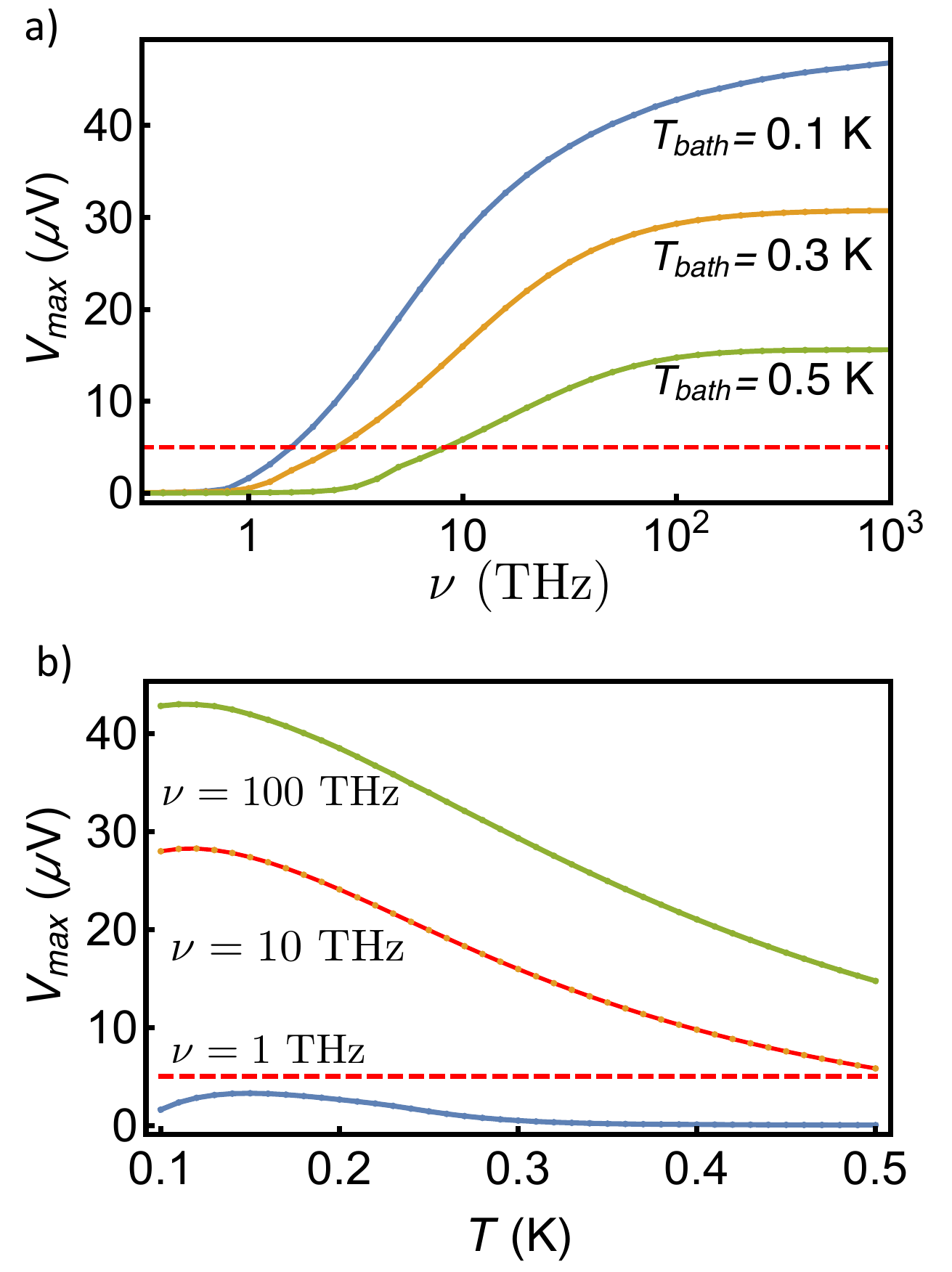}  
    \caption{ 
	 %{\bf Photon arrival discrimination.} 
	 a) The maximum voltage $V_{max}$ generated at the extremes of the SQUID as a function of the frequency of the absorbed photon at different bath  temperatures.
	b) $V_{max}$ as a function of the working  bath temperature for different energies of the incoming photons.
	Red dashed lines indicate a voltage threshold detection of $5~\mu$V.
     }  
    \label{fig:Freq_discrimination}
\end{figure} 

\section{Photon detection}

\subsection{Photon arrival detection}

The PSRD can be efficiently used to detect an arriving photon.
We can think to connect the device to a discriminator circuit that is triggered as soon as the voltage generated by the \DetectorName~exceeds a detection threshold $V_{min}$.
Figure \ref{fig:Freq_discrimination} a) shows the maximum voltage $V_{max}$ reached as a function of the photon frequency for different bath temperatures $T_{bath}=0.1, 0.3~$ and $0.5~$K.
If we set, for instance, a detection threshold of  $V_{min} = 5~\mu$V (red dashed line in Fig. \ref{fig:Freq_discrimination}), photons with  frequency $\nu$ above 10THz can be detected even at a bath temperature of $T_{bath} = 0.5$K.

Figure \ref{fig:Freq_discrimination} b) shows the maximum voltage $V_{max}$ as a function of bath temperature and for different photon frequencies.
For $V_{min} = 5~\mu$V, while photons of $1~$THz are at the limit of measurability, for higher frequency, i.e., $\nu \geq 10~$THz, the \DetectorName~is able to detect their arrivals even at high temperature.

\subsection{Photon frequency detection}

With the \DetectorName~it is also possible to determine the frequency of the detected photon, i.e., one can use it as a calorimeter.
In order to have a sizable voltage output we focus on photons with $\nu \geq 1~$THz.
To see if it is possible to distinguish the frequency of the absorbed photon, we need analyze the signal-to-noise (S/N) ratio and the resolving power.
The first can be defined as \cite{virtanen2017josephson}
\begin{equation}
  \frac{{\rm S}}{{\rm N}} (h \nu)= \frac{V(\nu=0, T_{bath}) - V( h \nu, T_{bath})}{ \sqrt{\mathcal{S}_V(T_{bath}) \omega}},
\end{equation}
where $V(\nu=0,T_{bath})$ and $V( h \nu, T_{bath})$ are the generated voltages when no-photon is absorbed and a photon of frequency $\nu$ is absorbed, respectively.
The function $\mathcal{S}_V(T_{bath})$ and $\omega$ are the voltage noise spectral density and the detector bandwidth, respectively.
The noise spectral density is determined by the Johnson noise spectral density  $\mathcal{S}_V(T_{bath}) = 4 k_B T_{bath} R_N$ in the absence of radiation.
The bandwidth must satisfy the relation $\omega \geq 2 \pi / \tau_{e-ph}$ \cite{virtanen2017josephson}.
Since we are working at moderately low temperature, i.e., $T_{bath} = 0.1~$K, we can assume  \cite{giazotto2008Ultrasensitive} $\tau_{e-ph} \approx 10^{-4}~$s.
Moreover, the  voltage pulse has typical width of a fraction of ns (see Fig. \ref{fig:RCSJ_dyn}) so that
to be able to detect it, we need a sufficiently high frequency resolution. 
In the following we consider a detector bandwidth of $\omega = 10~$GHz that should allow one to detect the main features of a large range of voltage pulses.
 
 \begin{figure}
  \includegraphics[scale=.6]{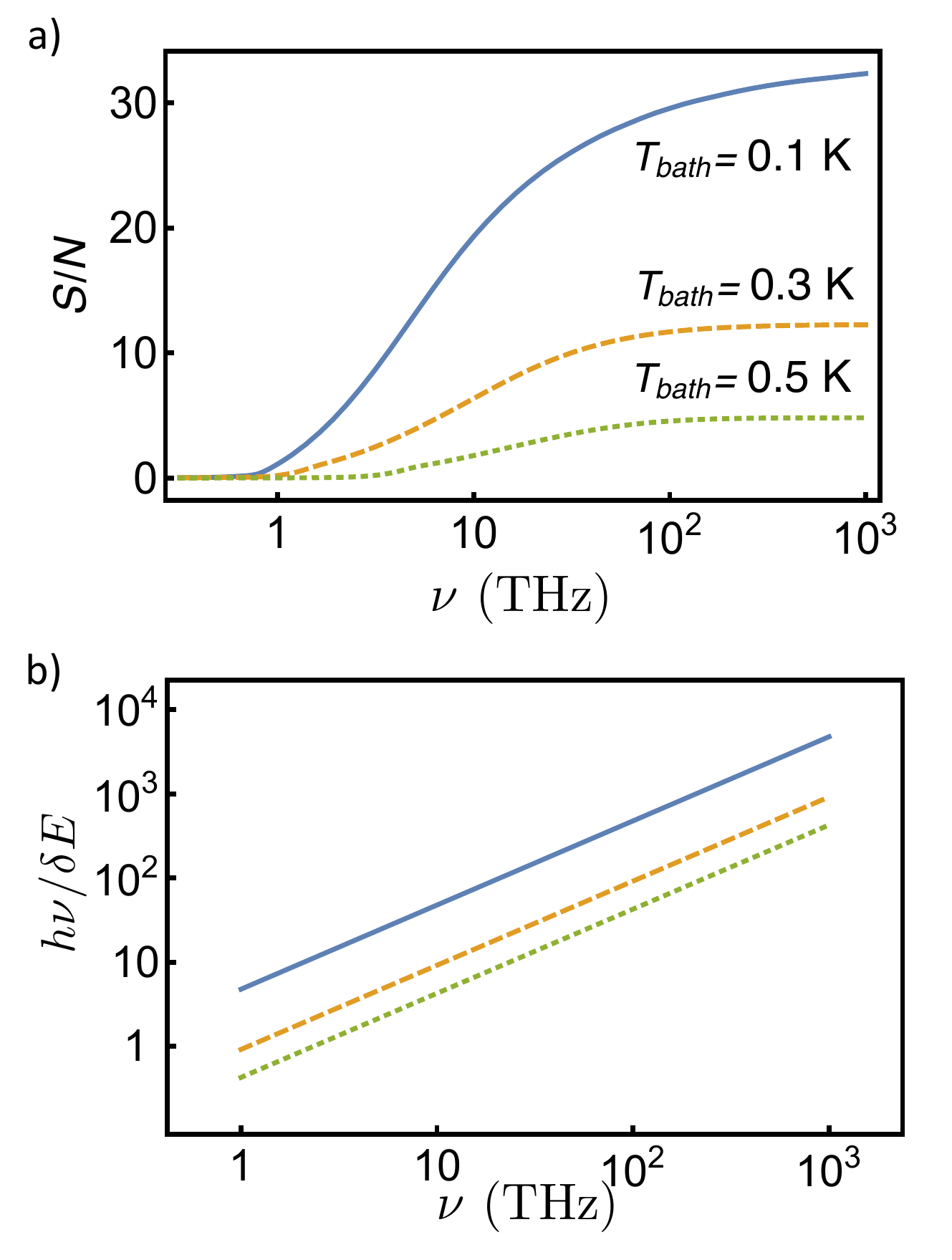}  
    \caption{ 
	 %{\bf Detector performances} 
	 a) PSRD signal-to-noise (S/N) ratio as a function of the frequency of the absorbed photon, and for different bath temperatures. The detection bandwidth is set to $\omega = 10~$GHz.
	b) Resolving power $h\nu/\delta E$ as a function of the frequency of the absorbed photon, and for different bath temperatures.
	The temperatures are $T_{bath} =0.1, 0.3, 0.5~$K (solid blue, dashed orange, and  dotted green line, respectively) as in the previous panel a).	
     }  
    \label{fig:SN_ResPow}
\end{figure} 

\begin{figure}
  \includegraphics[scale=.6]{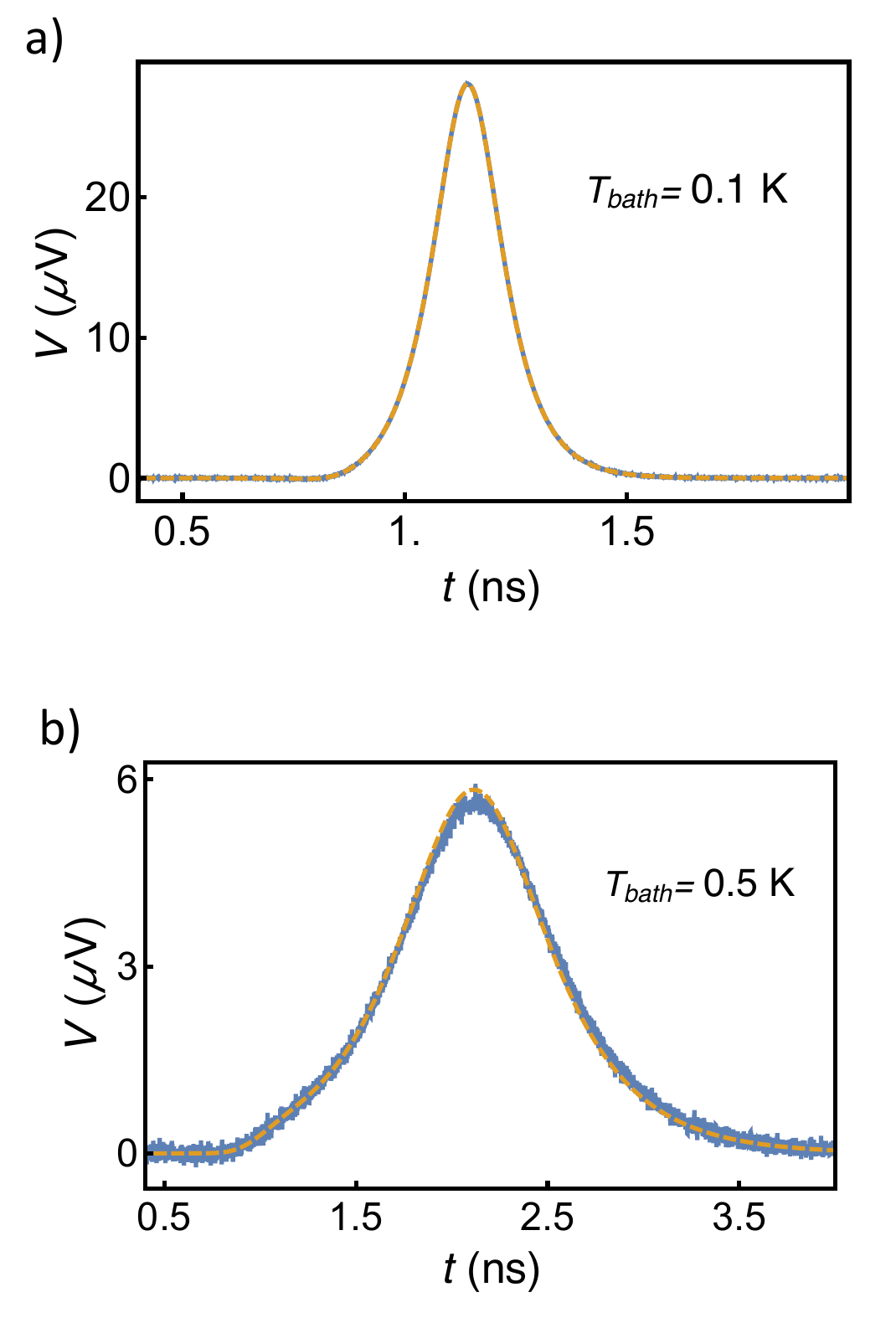}  
    \caption{ 
	 %{\bf Detector performances in presence of thermal noise} 
	 Comparison between the noisy (blue curve) and noiseless voltage dynamics (dashed yellow curve) for different bath temperatures:
	 a) $T_{bath}=0.1~$K and b) $T_{bath}=0.5~$K. The dynamics is obtained by averaging over $500$ noise dynamical realizations. For $T_{bath}=0.1~$K the effect of the noise is barely visible. The other chosen parameters are $\epsilon = 0.98$, and $\Phi/\Phi_0 = 0.499$.
     }  
    \label{fig:FigLangevin_dyn}
\end{figure}

With the above assumption, we get for the S/N ratio
\begin{equation}
 \Big | \frac{{\rm S}}{{\rm N}} (h \nu) \Big |=  \frac{V_{max}( h \nu, T_{bath})}{ \sqrt{ 4 k_B T_{bath} R_N \omega}},
\end{equation}
where $V_{max}( h \nu, T_{bath})$ is the maximum voltage generated by the photon, and extracted from Fig. \ref{fig:Freq_discrimination} a).
The behavior of the ${\rm S}/\rm N$ ratio for  $\omega = 10~$GHz is shown in Fig. \ref{fig:SN_ResPow} a) and, as expected,  closely resembles the shape of the maximum achievable voltage displayed in Fig. \ref{fig:Freq_discrimination} a).
We notice that, for a photon with $\nu \geq 10~$THz, we obtain a large ${\rm S}/\rm N \geq 5$ even at a high working temperature $T_{bath} =0.5~$K.
In the lower frequency range, the radiation sensing limitation is given by the voltage signal strength as discussed in the previous section.

The other figure of merit that characterizes the PSRD performance is the resolving power defined as \cite{virtanen2017josephson}
\begin{equation}
 \frac{h \nu}{\delta E} = \frac{ h \nu}{4 \sqrt{ 2 \log 2 k_B T^2_{bath} C_e   } }.
\end{equation}
As shown in Fig. \ref{fig:SN_ResPow} b), $h \nu/ \delta E$ increases linearly with the photon frequency and for $\nu \geq 10~$THz exceeds $10$ at $T_{bath} = 0.1~$K.

\section{Thermal noise effect}
\label{sec:thermal_noise}

To achieve good detecting performance the \DetectorName~must be set in a unstable state (determined by the externally applied magnetic flux).
The photon arrival induces a perturbation and thereby the phase dynamics and the voltage generation across the SQUID.
In principle, any perturbation can induce the phase dynamics.
For this reason, it is of crucial importance to analyze the impact of external noise sources.

In the following we consider the effect of the thermal environment which is expected to be the predominant source of noise.
The dynamics of the \DetectorName~is determined by a modified RCSJ equation that includes a thermal noise contribution.
This can be written in terms of a Langevin RCSJ equation \cite{Solinas2015JRC}
\begin{equation}
	 \frac{\hbar}{2 e R} \dot{\varphi} + \alpha[ f(\varphi,\tau) - \delta] =  \xi(t),
	\label{eq:Langevin}
\end{equation}
where $ \xi(t)$ is the white noise contribution with correlation function
\begin{equation}
 \langle \xi(t) \xi(t') \rangle = \frac{2 k_B T_{bath}}{R_N} \delta(t-t').
\end{equation} 

The voltage dynamics in the presence of noise averaged over $500$ noise realizations is shown in Fig. \ref{fig:FigLangevin_dyn}.
As it can be seen, the effect of noise is barely visible up to high temperatures with $T_{bath}=0.5~$K.
Furthermore, even in this situation, the main dynamical features such as the shape and the maximum amplpitude of the voltage pulse are fully preserved.
From this follows that the \DetectorName~performance is almost unmodified by the presence of thermal noise.

\section{Conclusions}

In summary, we have proposed the original design for a single-photon sensor based on the proximity effect, and on the instability induced by the absorption of a photon.
In a SQUID configuration, two proximity superconductor-normal metal-superconductor junctions can be set in an unstable state by controlling an applied magnetic flux piercing the superconducting loop.
The absorbed photon induces an enhancement of the electronic temperature in the absorbing junction and, as final result, a voltage spike across the interferometer.
With our choice of realistic parameters for the structure, this signal is strong enough to be measured with conventional electronics. 
The voltage pulse shape and its amplitude depend on the photon frequency, the junction asymmetry and working bath temperature, and could be measured in several different configurations as soon as the photon frequency is larger than $\sim 1~$THz.
A straightforward use of the PSRD is as a photon arrival detector.
In addition, it can also work as a calorimeter since it is able to distinguish the absorbed photon frequency in the range $1~$THz$\leq \nu \leq 10^3~$THz.
The achievable signal-to-noise ratio and resolving power suggest a high performance when the PSRD works at moderately low temperatures around $0.1~$K.
Furthermore, the detector performance are basically unaffected by the working temperature (between $0.1~$K and $0.5~$K) and the presence of thermal noise.
These features make the PSRD interesting for a number of different applications, for instance, 
we can envision its use in quantum technologies as well as in astrophysics.

%\section{GHz region photons}
%
%To detect lower frequency photon, we need to change the detector characteristics.
%In fact, it is not enough to work at lower temperature since the induced asymmetry is decreased.
%Thus, we need to change also the junction length and the magnetic flux working point.
%
%In Fig. \ref{fig:Fig_GHz}, we show some result for incoming photons in the GHz region.
%These are obtained for a strong asymmetric SQUID with $\epsilon=l_1/l_2=0.5$.
%This limit the working temperature since we must satisfy the long junction limit $E_{Th}/k_B < T_e$.
%
%As we can see, with a parameter optimization we can detect photons with $\nu =100~$GHz working at $T_{bath}=0.3~$K [Fig. \ref{fig:Fig_GHz} a)].
%For lower frequencies, the voltage maximum decreases and is about $1.5~\mu$V for $\nu =10~$GHz  [Fig. \ref{fig:Fig_GHz} b)].
%
%
%\begin{figure}
%  \includegraphics[scale=.7]{Fig_GHz.pdf}  
%    \caption{ 
%	 {\bf Detector performances} a) The detector signal-to-noise  as a function of the frequency of the absorbed photon and for different temperatures.
%	b) The resolving power as a function of the frequency of the absorbed photon and for different temperatures.
%	Temperature are taken $T_{bath} =0.1, 0.3, 0.5~$K as in the previous figure \ref{fig:Freq_discrimination} a).	
%     }  
%    \label{fig:Fig_GHz}
%\end{figure} 
%

%%%%%%%%%%%%%%%%%%%%%%%%%%%%%%%%%%%%%%%%%%%%%%%%%%%%
\begin{acknowledgments}
PS has received funding from the European Union FP7/2007-2013 under REA
grant agreement no 630925 -- COHEAT and from MIUR-FIRB2013 -- Project Coca (Grant
No.~RBFR1379UX). 
FG has eceived funding from the European Research Council under the European Union’s Seventh Framework Program (FP7/2007-2013)/ERC Grant agreement No. 615187-COMANCHE, and Tuscany Region under the FARFAS 2014 project SCIADRO.
GPP  e PS have received funding from the Progetto Premiale Q-SecGroundSpace--
Intermodal Secure Quantum Communication on Ground and Space D.M. n.543 del 05/08/2015.
\end{acknowledgments}

%%%%%%%%%%%%%%%%%%%%%%%%%%%%%%%%%%%%%%%%%%%%%%%%%%%%
%\bibliographystyle{apsrev4-1}
%
%\bibliographystyle{apsrev4-1}
%\bibliography{bibQPJ}
%%%%%%%%%%%%%%%%%%%%%%%%%%%%%%%%%%%%%%%%%%%%%%%%%%%%
%

%%%%%%%%%%%%%%%%%%%%%%%%%%%%%%%%%%%%%%%%%%%%%%%%%%%%%%%
\end{document}